\documentclass[aps,pra,twocolumn,amsmath,amssymb,showpacs,10pt]{revtex4-1}
\usepackage{times}

\usepackage[english]{babel}
\usepackage{graphicx}
\usepackage{dcolumn}
\usepackage{bm}
\usepackage{color}
\usepackage{amsmath}
\usepackage{relsize,amsmath,dsfont,mathrsfs,empheq,verbatim,upgreek}
\usepackage{etoolbox}
\usepackage[caption = false]{subfig}
\apptocmd{\thebibliography}{\raggedright}{}{}

\newcommand{\tmop}[1]{\ensuremath{\operatorname{#1}}}

\newcommand{\mathd}{\mathrm{d}}

\newcommand{\bra}[1]{\langle #1 |}
\newcommand{\ket}[1]{| #1 \rangle}

\newcommand{\mean}[1]{\langle #1 \rangle}

\newcommand{\norm}[1]{\| #1 \|}
\newcommand{\Id}{\mathds{1}}

\begin{document}
\newdimen\origiwspc%
  \newdimen\origiwstr%

\title{Quantum regression theorem and non-Markovianity of quantum dynamics}

\author{Giacomo Guarnieri $^{1,2}$, Andrea Smirne $^{3,4}$, Bassano Vacchini $^{1,2}$}

\affiliation{$^1$\mbox{Dipartimento di Fisica, Universit{\`a} degli Studi di Milano, Via Celoria 16, 20133 Milan, Italy}\\
$^2$\mbox{Istituto Nazionale di Fisica Nucleare, Sezione di Milano, Via Celoria 16, 20133 Milan, Italy}\\
$^3$\mbox{Dipartimento di Fisica, Universit{\`a} degli Studi di Trieste, Strada Costiera 11, 34151 Trieste, Italy} \\
$^4$\mbox{Istituto Nazionale di Fisica Nucleare, Sezione di Trieste, Via Valerio 2, 34127 Trieste, Italy}}

\begin{abstract}
  We explore the connection between two recently introduced notions of
  non-Markovian quantum dynamics and the validity of the so-called
  quantum regression theorem. While non-Markovianity of a quantum
  dynamics has been defined looking at the behaviour in time of the
  statistical operator, which determines the evolution of mean values,
  the quantum regression theorem makes statements about the behaviour
  of system correlation functions of order two and higher. The
  comparison relies on an estimate of the validity of the quantum
  regression hypothesis, which can be obtained exactly evaluating two
  points correlation functions. To this aim we consider a qubit
  undergoing dephasing due to interaction with a bosonic bath,
  comparing the exact evaluation of the non-Markovianity measures with
  the violation of the quantum regression theorem for a class of
  spectral densities. We further study a photonic dephasing model,
  recently exploited for the experimental measurement of
  non-Markovianity.  It appears that while a non-Markovian dynamics
  according to either definition brings with itself violation of the
  regression hypothesis, even Markovian dynamics can lead to a failure
  of the regression relation.
\end{abstract}
 
\date{\today}
\maketitle

\section{Introduction}
In recent times there has been a revival in the study of the
characterization of non-Markovianity for an open quantum system
dynamics.  While the subject was naturally born together with the
introduction of the first milestones in the description of the time
evolution of a quantum system interacting with an environment
\cite{Lindblad1976a,Lindblad1979a}, the difficulty inherent in the
treatment led to very few general results, and the very definition of
a convenient notion of Markovian open quantum dynamics was not agreed
upon. The focus initially was on finding the closest quantum
counterpart of the classical notion of Markovianity for a stochastic
process, so that reference was made to correlation functions of all
order for the process. Recent work was rather focused on proposals of
a notion of Markovian quantum dynamics based on an analysis of the
behaviour of the statistical operator describing the system of interest
only, thus concentrating on features of the dynamical evolution map,
which only determines mean values. Different properties of the time
evolution map have been considered in this respect \cite{Wolf2008PRL,Breuer2009PRL,Breuer2012JPB,Rivas2010PRL,Lu2010PRA,LuoPRA2012,Lorenzo2013PRA,Bylicka2013arxiv,Chruscinski2014PRL}.
In particular two viewpoints \cite{Breuer2009PRL,Rivas2010PRL} appear
to have captured important aspects in the characterization of a
dynamics which can be termed non-Markovian in the sense that it
relates to memory effects.

The aim of our work is to analyse the relationship between these
approaches and the validity of the so-called quantum regression theorem
\cite{Breuer2002,Gardiner2004}, according to which the behaviour in
time of higher order correlation functions can be predicted building
on the knowledge of the dynamics of the mean values for a generic
observable. The analysis can be performed introducing a suitable
quantifier for the violation of the quantum regression hypothesis,
which in turn requires knowledge of the exact two-time correlation
functions. We therefore consider a two-level system coupled to a
bosonic bath through a decoherence interaction, exactly estimating for a
general class of spectral densities the predictions of different
criteria for non-Markovianity of a dynamics and the violation of the
regression theorem. We further apply this analysis to a dephasing
model, whose realization has been recently exploited to experimentally
observe quantum non-Markovianity \cite{Liu2011NAT}. In both cases we
show that the quantum regression theorem can be violated even in the
presence of a quantum dynamics which, according to either criteria, is
considered Markovian.

The paper is organized as follows. In Sect.~\ref{sec:nmarkov} we
recall two recently introduced notions of Markovianity for a quantum
dynamics and the associated measures, while in Sect.~\ref{sec:qrt} we
address the formulation of the quantum regression theorem and
introduce a simple estimator for its violation. We apply this
formalism to the pure dephasing spin Boson model in Sect.~\ref{sec:sbm}
discussing the relationship between the two approaches, and extend the
analysis to a photonic dephasing model in
Sect.~\ref{sec:modellofotoni}. We finally comment on our results in Sect.~\ref{sec:ceo}.

\section{Non-Markovianity definitions and measures}
\label{sec:nmarkov}

Let us start by briefly recalling the main features of the notion of non-Markovian
quantum dynamics which will be exploited in the following analysis.
In the classical theory of stochastic processes, the
definition of Markov process involves 
the entire hierarchy of $n-$point joint probability distributions
associated with the process. Since such a definition
cannot be directly transposed to the quantum realm \cite{Vacchini2011NJP,Logullo2014arxiv},
different and non-equivalent notions of quantum Markovianity have been
introduced \cite{Wolf2008PRL,Breuer2009PRL,Breuer2012JPB,Rivas2010PRL,Lu2010PRA,LuoPRA2012,Lorenzo2013PRA,Bylicka2013arxiv,Chruscinski2014PRL},
along with different measures {to quantify the degree of non-Markovianity of a given dynamics}
(see \cite{Addis2014arxiv,Rivas2014arxiv} for a very recent comparison).
These definitions {all} convey the idea that the occurrence of memory effects is the proper attribute of non-Markovian dynamics,
relying on different properties of the dynamical maps
which describe the evolution of the open quantum system.
In the absence of initial correlations between the open system and its environment, i.e.,
\begin{equation}\label{eq:prod}
\rho_{SE}(0) = \rho_S(0) \otimes \rho_E(0)
\end{equation}
{with $\rho_E(0)$ assumed to be fixed}, the evolution of an open quantum system
is {characterized} by a one parameter family of completely positive and trace preserving (CPT) maps $\left\{\Lambda(t)\right\}_{t\geq 0}$, 
such that \cite{Breuer2002}
\begin{equation}\label{eq:red}
\rho_S(t) = \Lambda(t) \rho_S, 
\end{equation}
where $\rho_S \equiv \rho_S(0)$ is the state of the open system at the initial time $t_0=0$.
A relevant class of open quantum system's dynamics is provided by the semigroup ones, 
which are characterized by the composition law
\begin{equation}
\Lambda(t)\Lambda(s) = \Lambda(t+s) \qquad \forall t,s \geq 0.
\end{equation}
The generator of a semigroup of CPT maps is fixed by the Gorini-Kossakowski-Sudarshan-Lindblad
theorem \cite{Lindblad1976a,Gorini1976a},
which implies that the dynamics
of the system is given by the Lindblad equation
\begin{multline}
\frac{\mathd}{\mathd t} \rho_S(t) = - i[H, \rho_S(t)] \\
+ \sum_{k} \gamma_k\left(L_k \rho_S(t) L^{\dag}_k - \frac{1}{2} \left\{L^{\dag}_kL_k , \rho_S(t) \right\} \right) 
\end{multline}
with $\gamma_k \geq0$. 
The semigroups of CPT maps are identified with the Markovian time-homogeneous dynamics
according to all the previously mentioned definitions of Markovianity, so that the differences between them
actually concern the notion of time-inhomogeneous Markovian dynamics.

In the following, we will take into account two definitions of Markovianity and the corresponding measures of non-Markovianity.
One definition \cite{Breuer2009PRL} is related with the contractivity of the trace distance under the action of the dynamical maps,
while the other \cite{Rivas2010PRL} relies on a divisibility property of the dynamical maps, which reduces to the semigroup composition
law in the time-homogeneous case.

\subsection{Trace-Distance measure}\label{sec:nmblp}

The basic idea behind the definition of non-Markovianity introduced by Breuer, Laine and Piilo (BLP) \cite{Breuer2009PRL} 
is that a change in the distinguishability between the reduced states can be read in terms of an information flow between the open system and the environment.
The distinguishability {between} quantum states is quantified through the trace distance \cite{Nielsen2000},
which is the metric on the space of states induced by the trace norm: 
\begin{equation}\label{eq:tracedistance}
D(\rho^1, \rho^2) = \frac{1}{2} \norm{\rho^1-\rho^2}_1 = \frac{1}{2} \sum_k |x_k|,
\end{equation}
where the $x_k$ are the eigenvalues of the traceless hermitian operator $\rho^1-\rho^2$.
The trace distance takes values between 0 and 1 and, most importantly,
it is a contraction under the action of CPT maps. By investigating the evolution of the trace
distance between two states of the open system {coupled to the same environment but evolved} from different initial conditions,
\begin{equation}
D(t, \rho_S^{1,2}) \equiv D(\rho^1_S(t), \rho^2_S(t)),\quad \rho^k_S(t) = \Lambda(t) \rho^k_S,
\end{equation}
one can thus describe the exchange of information between the open system and the environment.
A decrease of the trace distance $D(t, \rho_S^{1,2})$ means a lower ability to discriminate
between the two initial conditions $\rho_S^{1}$ and $\rho^2_S$, which can be expressed by saying
that some information has flown out of the open system. On the same ground, an increase of the trace distance
can be ascribed to a back-flow of information to the open system
and then represents a memory effect in its evolution.
Non-Markovian quantum dynamics can be thus defined as those dynamics which present
a non-monotonic behaviour of the trace distance, i.e. such that there
are time intervals
$ \Omega_+ $ in which 
\begin{equation}\label{BLPsignature}
\sigma(t,\rho_S^{1,2}) = \frac{d}{dt} D(t,\rho_S^{1,2}) >0.
\end{equation}
Consequently, the non-Markovianity of an open quantum system's dynamics $\left\{\Lambda(t)\right\}_{t\geq 0}$ is quantified by the measure
\begin{equation}\label{eq:nblp}
\mathscr{N}= \max_{\rho_S^{1,2}}\int_{\Omega_+}\sigma(t,\rho_S^{1,2}) \, \mathd t.
\end{equation} 
The maximization involved in the definition of this measure
can be greatly simplified since
the optimal states
must be orthogonal \cite{Wissmann2012PRA} and, even more, one can determine
$\mathscr{N}$ by means of a local maximization over one state only \cite{Liu2014arxiv}.
This measure of non-Markovianity has been also investigated
experimentally in all-optical settings \cite{Liu2011NAT,Tang2012EPL,Liu2013SCI}.

\subsection{Divisibility measure}\label{sec:nmrhp}

The definition given by Rivas, Huelga and Plenio (RHP) \cite{Rivas2010PRL} identifies Markovian dynamics with those dynamics which are described by a CP-divisible family of quantum dynamical maps $ \lbrace \Lambda(t)\rbrace_{t\geq 0} $  (CP standing for completely positive), i.e. such that
\begin{equation} \label{divisibility}
\Lambda(t_2) = \Lambda(t_2,t_1)\Lambda(t_1) \qquad \forall \, t_2 \geq t_1 \geq 0,
\end{equation}
$\Lambda(t_2,t_1)$ being itself a completely positive map. 
Indeed, if $\Lambda(t_2,t_1) = \Lambda(t_2-t_1)$
the composition law in Eq.(\ref{divisibility}) is equivalent to the semigroup composition law.
An important property of this definition is that, provided that the evolution
of the reduced state can be formulated by a time-local master equation
\begin{eqnarray}
\label{Lindbladt}
\frac{\mathd}{\mathd t} \rho_S(t) 
&=& \mathcal{K}(t)[\rho_S(t)]\\ 
&=& - i[H(t), \rho_S(t)]
\nonumber
\\
&&
\hspace{-2truecm}
+
\sum_{k} \gamma_k(t)\left(L_k(t) \rho_S(t) L^{\dag}_k(t) -
  \frac{1}{2} \left\{L^{\dag}_k(t)L_k(t) , \rho_S(t) \right\} \right), 
\nonumber
\end{eqnarray}
the positivity of the coefficients,  $\gamma_k(t) \geq 0$ for any $t\geq 0$, is equivalent to the CP-divisibility
of the corresponding dynamics. This can be shown by taking into account the family of propagators $\Lambda(t_2,t_1)$
associated with Eq.(\ref{Lindbladt}),
\begin{equation}\label{eq:prop}
\Lambda(t_2,t_1) = T_{\leftarrow} \exp\left(\int^{t_2}_{t_1}\mathcal{K}(s) \mathd s\right),
\end{equation}
where $T_{\leftarrow}$ denotes the time ordering and $\Lambda(t,0) \equiv \Lambda(t)$.
By construction, the propagators $\Lambda(t_2,t_1)$ satisfy Eq.(\ref{divisibility}),
but, in general, they are not CP maps. One can show \cite{Laine2010PRA,Rivas2012} that the propagators are actually CP
if and only if the coefficients $\gamma_k(t)$ are positive functions of time.

The corresponding measure of non-Markovianity is given by
\begin{equation}
\mathcal{I} = \int_{\mathbb{R}^+} \, dt \, \mathfrak{g}(t)
\end{equation}
with
\begin{equation}\label{eq:RHPMeasure}
\mathfrak{g}(t) = \lim_{\epsilon\to 0^+} \frac{\frac{1}{N}\norm{\Lambda_{Choi}(t,t+\epsilon)}_1-1}{\epsilon},
\end{equation}
where $ \Lambda_{Choi} $ is the Choi matrix associated with $\Lambda$.
Given a maximally entangled state between the system and an ancilla, $ \ket{\psi} = \frac{1}{\sqrt{N}} \sum_{k=1}^N \ket{u_k} \otimes \ket{u_k} $, 
one has \cite{Choi1975}
\begin{align}\label{Choi}
\Lambda_{Choi} = N \left(\Lambda\otimes\Id_N\right)\left( \ket{\psi}\bra{\psi} \right).
\end{align}
The positivity of the Choi matrix corresponds to the complete positivity of the map $\Lambda$ and it is equivalent to the condition $\|{\Lambda_{Choi}}\|_1=N$,
so that the quantity $ \mathfrak{g}(t) $ is different from zero if and only if the CP-divisibility of the dynamics is broken.

Finally, since the trace distance is contractive under CPT maps, if a dynamics is Markovian according to the RHP definition,
then it is so also according to the BLP definition, i.e.,
\begin{equation}\label{eq:impl}
\mathcal{I} = 0 \Longrightarrow \mathscr{N} = 0,
\end{equation}
while the opposite implication does not hold \cite{Laine2010PRA,Haikka2011PRA,Chruscinski2011PRA}.

\section{The quantum regression theorem} \label{sec:qrt}
As recalled in the Introduction, the quantum regression theorem provides a benchmark structure
in order to study the multi-time correlation functions of an open quantum system.
For the sake of simplicity, we {focus} on the two-time correlation functions only.
Given two open system's operators, $A \otimes \Id_E$ and $B \otimes \Id_E$,
where $\Id_E$ denotes the identity on the Hilbert space associated with the environment,
their two-time correlation function is defined as  
\begin{multline}\label{twotimecorrfunc}
\langle A(t_2) B(t_1) \rangle \equiv \tmop{Tr} \left[ U^{\dagger}(t_2) A\otimes \Id_E U(t_2) \right.\\
\left. 
\times
 U^{\dagger}(t_1) \left( B\otimes \Id_E\right) U(t_1)\rho_{SE}(0)\right],
\end{multline}
where $U(t)$ is the overall unitary evolution operator and we set $t_2 \geq t_1 \geq 0$. In the following,
we assume an initial state as in Eq.(\ref{eq:prod}), as well as a time-independent total Hamiltonian
$H_T = H_S \otimes \Id_E + \Id_S \otimes H_E + H_I$, so that $U(t) = e^{-i H_T t}$.

The condition of an initial product state with a fixed environmental
state guarantees the existence of a reduced dynamics, see
Eqs.(\ref{eq:prod}) and (\ref{eq:red}).  This means that all the
one-time probabilities associated with the observables of the open
systems and, as a consequence, their mean values can be evaluated by
means of the family of reduced dynamical maps only, without need
for any further reference to the overall unitary dynamics.  An
analogous result holds for the two-time correlation functions, if one
can apply the so-called quantum regression theorem.  The latter
essentially states that under proper conditions the dynamics of the
two-time correlation functions can be reconstructed from the dynamics
of the mean values, or, equivalently, of the statistical operator.
Indeed, if the quantum regression theorem cannot be applied, one needs
to come back to the full unitary dynamics in order to determine the
evolution of the two-time correlation functions.
We will not repeat here the detailed derivation of the quantum
regression theorem, which can be found in \cite{Carmichael1993,Breuer2002,Gardiner2004}.
Nevertheless, let us recall the basic ideas.
First, by introducing the operator
\begin{equation}
\chi(t_2,t_1) = e^{-i H_T(t_2-t_1)} B \otimes \Id_E \rho_{SE}(t_1) e^{i H_T(t_2-t_1)},
\end{equation}
the two-time correlation function in Eq.(\ref{twotimecorrfunc}) can be rewritten as
\begin{equation}\label{eq:aux}
\langle A(t_2) B(t_1) \rangle = \tmop{Tr}_S{A \tmop{Tr}_E{ \chi(t_2,t_1)}}.
\end{equation}
Now, suppose that we can describe the evolution of $\chi(t_2,t_1)$
with respect to $t_2$ with the same dynamical maps which fix the evolution of the statistical operator, i.e.,
\begin{equation}\label{eq:aux2}
\chi(t_2, t_1) = \Lambda(t_2,t_1)[ \chi(t_1,t_1)],
\end{equation} 
where $\Lambda(t_2,t_1)$ is the propagator introduced in Eq.(\ref{eq:prop}).
Then, Eq.(\ref{eq:aux}) directly provides
\begin{equation}\label{eq:preqrt}
\langle A(t_2) B(t_1) \rangle_{qrt} = \tmop{Tr}_S{A \, \Lambda(t_2,t_1)[ B \rho_S(t_1)]}.
\end{equation}
The two-time correlation functions can be fully determined by the dynamical
maps which fix the evolution of the statistical operator: the validity of Eq.(\ref{eq:preqrt})
can be identified with the validity of the quantum regression theorem and we
will use the subscript $qrt$ to denote the two-time correlation functions evaluated through Eq.(\ref{eq:preqrt}).
Indeed, all the procedure relies on Eq.(\ref{eq:aux2}), which requires that the same assumptions
made in order to derive the dynamics of $\rho_S(t)$ can be made also
to get the evolution of $\chi(t_2,t_1)$ with respect to $t_2$ \cite{Gardiner2004}.
Especially, the hypothesis of an initial total product state in Eq.(\ref{eq:prod})
turns into the hypothesis of a product state at any intermediate time $t_1$,
\begin{equation} \label{eq:prodt}
\rho_{SE}(t_1) = \rho_S(t_1) \otimes \rho_E.
\end{equation}
The physical idea is that the quantum regression theorem holds when
the system-environment correlations due to the interaction can be neglected \cite{Swain1981}.
Note that this condition will never be strictly satisfied, as long as the system
and the environment mutually interact, but it should be understood as a guideline
to detect the regimes in which Eq.(\ref{eq:preqrt}) provides a satisfying description of the evolution of the two-time correlation functions.
More precisely, D{\"u}mcke \cite{Dumcke1983} demonstrated that the exact expression of the two-time (multi-time) correlation
functions, see Eq.(\ref{twotimecorrfunc}), converges to the expression in Eq.(\ref{eq:preqrt}) in the weak coupling limit and in the singular coupling limit. 
As well-known,
in these limits the reduced dynamics converges to a semigroup dynamics \cite{Davies1974,Gorini1976}.
Nevertheless, the correctness of a semigroup description of the reduced
dynamics is not always enough to
guarantee the accuracy of the quantum regression theorem \cite{Talkner1986,Ford1996}.
More in general, 
the precise link between a sharply defined notion of Markovianity of quantum dynamics
and the quantum regression theorem has still to be investigated.

The quantum regression theorem provided by Eq.(\ref{eq:preqrt}) can be 
equivalently formulated in terms of
the differential equations satisfied by mean values
and two-time correlation functions, as was originally done in \cite{Lax1968}.
For the sake of simplicity, let us restrict to the finite dimensional case, i.e.,
the Hilbert space associated with the open system is $\mathbb{C}^N$.
Consider a reduced dynamics fixed by the family of maps $\left\{\Lambda(t)\right\}_{t\geq0}$
and a basis $ \lbrace E_i \rbrace_{1, \ldots, N^2}$ of linear operators on $\mathbb{C}^N$,
such that the corresponding mean values fulfill the coupled linear equations of motion \cite{Carmichael1993}
\begin{equation}\label{lindiffeq}
\frac{d}{dt} \mean{E_i(t)} = \sum_j G_{ij} (t) \mean{E_j(t)},
\end{equation}
with the initial condition $\mean{E_i(t)}|_{t=0} = \mean{E_i(0)}$.
In this case, the quantum regression theorem is said to hold if the two-time correlation functions satisfy \cite{Breuer2002,Gardiner2004}
\begin{equation}\label{qrtdiscrete}
\dfrac{d}{dt_2}\langle E_i(t_2) E_k (t_1) \rangle_{qrt} = \sum_j G_{ij}(t_2)\langle E_j(t_2) E_k (t_1) \rangle_{qrt}\,,
\end{equation}
with the initial condition 
$$
\langle E_i(t_2) E_k (t_1) \rangle |_{t_2=t_1} = \langle E_i(t_1) E_k (t_1) \rangle.
$$

In the following, we will compare the evolution of the exact two-time correlation functions
obtained from the full unitary evolution $\langle E_i(t_2) E_k(t_1) \rangle$, see Eq.(\ref{twotimecorrfunc}),
with those predicted by the quantum regression theorem $\langle E_i(t_2) E_k(t_1) \rangle_{qrt}$.
To quantify the error made by using the latter, we exploit the relative error, i.e., we use the following figure of merit:
\begin{equation}\label{eq:figuremerit}
Z \equiv \left| 1 - \frac{\langle A(t_2) B(t_1) \rangle_{qrt}}{\langle A(t_2) B(t_1) \rangle} \right|,
\end{equation}
which depends on the chosen couple of open system's operators.
Hence, in general, one should consider different estimators, one for each
couple of operators in the basis $ \lbrace E_i \rbrace_{1, \ldots, N^2}$, and a maximization
over them could be taken. Nevertheless, in the following analysis 
it will be enough to deal with a single couple of system's operators, 
which fully encloses the violations of the quantum regression
theorem for the models at hand.

\section{Pure-dephasing spin boson model}\label{sec:sbm}
In this section, we take into account a model whose full unitary evolution
can be exactly evaluated \cite{Unruh1995,Breuer2002}, so as
to obtain the exact expression of the two-time correlation functions, to be compared
with the expression provided by the quantum regression theorem.
This model is a pure-decoherence model, in which the decay
of the coherences occurs without a decay of
the corresponding populations. Indeed, this is due to the fact that the free
Hamiltonian of the open system $H_S \otimes \Id_E$ commutes with the total Hamiltonian $H_T$ \cite{Breuer2002}.

\subsection{The model}\label{sec:tm}

Let us consider a two-level system linearly interacting with a bath of harmonic oscillators, so that the total Hamiltonian is
\begin{equation}\label{eq:Hamiltonian}
H_{T}\!=\!\frac{\omega_s}{2}\sigma_z\otimes\Id_E+\Id_S\otimes\sum_k\omega_kb^{\dagger}_k b_k +\sum_k \sigma_z \otimes \left( g_k b^{\dagger}_k + g^*_k b_k \right)
\end{equation}
The unitary evolution operator of the overall system in the interaction picture is given by \cite{Breuer2002}
\begin{equation}
\label{eq:EvolutionOperator}
U(t) = e^{i \Psi(t)} V(t),
\end{equation}
where the first factor is an irrelevant global phase and the second factor is the unitary operator
\begin{equation}\label{eq:EvolutionOperatorParticularForm}
V(t) = \exp \left[ \frac{1}{2} \sigma_z \otimes  \sum_k \left( \alpha_k(t) b^{\dagger}_k- \alpha^*_k(t)  b_k \right) \right],
\end{equation}
with
\begin{equation}\label{eq:alpha}
\alpha_k(t) = 2 g_k \frac{1- e^{i \omega_k t}}{\omega_k}.
\end{equation}
The reduced dynamics is readily calculated to give
\begin{equation}
\label{rho_S(t)matrix}
\rho_S(t)= \begin{pmatrix}
\rho_{00} & \rho_{01}\gamma(t)e^{-i\omega_s t} \\
\rho_{10} \gamma^*(t)e^{i\omega_s t} & \rho_{11}
\end{pmatrix},
\end{equation}
where the function $\gamma(t)$ is given by
\begin{align}\label{decoherentfunction}
\gamma(t) &= \tmop{Tr}_E{\rho_E \prod_k \exp\left[\alpha_k(t) b^{\dagger}_k - \alpha^*_k(t) b_k\right]} \notag\\
&= \tmop{Tr}_E{\rho_E \prod_k \Delta(\alpha_k(t)) },
\end{align}
$ \Delta(\alpha) $ being the displacement operator of argument $\alpha$ \cite{Ferraro2005}.
The associated master equation reads
\begin{equation}\label{eq:me}
\frac{\mathd }{\mathd t} \rho_S(t) = -i \frac{\epsilon(t)}{2} [\sigma_z, \rho_S(t)] + \frac{\mathcal{D}(t)}{2} \left(\sigma_z \rho_S(t) \sigma_z - \rho_S(t) \right),
\end{equation}
where
\begin{equation}\label{eq:eps}
\epsilon(t) =  \omega_s -  {Im}\left[\frac{\mathd \gamma(t)/\mathd t}{\gamma(t)}\right]
\end{equation}
and the so-called dephasing function $\mathcal{D}(t)$ is 
\begin{equation}\label{eq:dt}
\mathcal{D}(t) = - {Re}\left[\frac{\mathd \gamma(t)/\mathd t}{\gamma(t)}\right] = -  \frac{\mathd}{\mathd t} \ln|\gamma(t)|.
\end{equation}
In the following, we will focus on the case of an initial thermal state of the bath, $\rho_E = \exp(-\beta H_E)/Z$
with $Z= \tmop{Tr}_E{ \exp(-\beta H_E)}$ and $\beta = (k_B T)^{-1}$ the inverse temperature.
We also consider the continuum limit: given
a frequency distribution $f(\omega)$ of the bath modes,
we introduce the
spectral density $J(\omega) = 4 f(\omega) |g(\omega)|^2$,
so that one has \cite{Breuer2002} 
\begin{equation}
\gamma(t)= \exp\left[ -\int_0^{\infty} d\omega \, J(\omega) \coth\left( \frac{\beta\omega}{2} \right) \frac{1-\cos\left(\omega t\right)}{\omega^2}\right],
\end{equation}
and hence $\epsilon(t)=\omega_s$ and
\begin{equation}\label{eq:Dephasing}
\mathcal{D}(t) = \int_0^{\infty} \,d\omega\,J(\omega) \coth\left( \frac{\beta\omega}{2} \right) \frac{\sin\left(\omega t\right)}{\omega}.
\end{equation}

\subsection{Measures of non-Markovianity}\label{sec:nmm}

\subsubsection{General expressions}

For this specific model, the two definitions of Markovianity are actually equivalent \cite{Zeng2011PRA}, i.e. not only Eq.\eqref{eq:impl} holds, but also the opposite does so.
{This is due to the fact that there is only one operator contribution in the time-local master equation \eqref{eq:me}, corresponding to the dephasing interaction.
}Nevertheless, the numerical values of the two measures of non-Markovianity are in general different and, more importantly, they depend 
in a different way on the parameters of the model.

Let us start by evaluating the BLP measure, see Sec. \ref{sec:nmblp}.
The trace distance between two reduced states evolved through Eq.(\ref{rho_S(t)matrix}) is given by
\begin{equation}
D(t,\rho_S^{1,2}) = \sqrt{\delta_p^2 + |\delta_c|^2 |\gamma(t)|^2},
\end{equation}
where  $ \delta_p= \rho^1_{00}-\rho^2_{00} $ and $ \delta_c= \rho^1_{01}-\rho^2_{01} $ are the differences between, 
respectively, the populations and the coherences of the two initial conditions $\rho^1_S$ and $\rho^2_S$. 
The couple of initial states that maximizes the growth of the trace distance 
is given by the pure orthogonal states 
$\rho_S^{1,2}= \ket{\psi_\pm}\bra{\psi_\pm} $, where $ \ket{\psi_\pm} = \frac{1}{\sqrt{2}}\left( \ket{0} \pm\ket{1} \right) $,
and the corresponding trace distance at time $t$ is simply $|\gamma(t)|$. 
The BLP measure therefore reads
\begin{equation}\label{eq:nmblp}
\mathscr{N} = \sum_m \left( |\gamma(b_m)| -|\gamma(a_m)| \right),
\end{equation}
where $ \Omega_+ = \bigcup_m \left( a_m,b_m\right) $ is the union of the time intervals in which $ |\gamma(t)|$ increases.
The BLP measure is different from zero if and only if $\mathd |\gamma(t)|/\mathd t > 0$ for some interval of time,
which is equivalent to the requirement that the dephasing function $\mathcal{D}(t)$ in Eq.(\ref{eq:me}) 
is not a positive function of time, i.e., that the CP-divisibility of the dynamics is broken, Sec.~\ref{sec:nmrhp}. 
As anticipated, for this model $\mathscr{N}>0 \Longleftrightarrow \mathscr{I} >0$.
Furthermore, given a pure dephasing master equation as in Eq.(\ref{eq:me}), one has \cite{Rivas2010PRL}
$\mathfrak{g}(t) = 0$ if $\mathcal{D}(t) \geq 0$ and $\mathfrak{g}(t) = - \mathcal{D}(t)$ if $\mathcal{D}(t) < 0$,
so that, see Eq.(\ref{eq:dt}),
\begin{equation}\label{eq:nmrhp}
\mathscr{I} = \sum_m \left( \ln|\gamma(b_m)| -\ln|\gamma(a_m)| \right), 
\end{equation}
where the $a_m$ and $b_m$ are defined as for the BLP measure. 

\subsubsection{Zero-temperature environment}\label{sec:zte}
In order to evaluate explicitly the non-Markovianity measures, we need to specify the
spectral density $J(\omega)$. In the following, we assume a spectral density of the form
\begin{equation}\label{spectralS}
J(\omega) = \lambda\frac{\omega^s}{\Omega^{s-1}}e^{-\frac{\omega}{\Omega}}
\end{equation} 
where $ \lambda $ is the coupling strength, the parameter $s$ fixes
the low frequency behaviour and $ \Omega $ is a cut-off frequency. 
The non-Markovianity for the pure dephasing spin model with a spectral density as in Eq.(\ref{spectralS}) has 
been considered in \cite{Haikka2013PRA,Addis2014arxiv} for the case
$\lambda = 1$.
We are now interested in the comparison between non-Markovianity and
violations of the quantum regression theorem, so that, as will become clear in the next section, the dependence on $\lambda$ plays a crucial role. In particular,
we consider the case of low temperature, i.e., $\beta \gg 1$,  so that $ \coth\left( \frac{\beta\omega}{2} \right)\approx 1 $.
The dephasing function in this case reads, see Eq.(\ref{eq:Dephasing}),
\begin{equation}\label{dephasingT0}
\mathcal{D}_s(t) = \frac{\lambda\Omega\Gamma(s)}{\left( 1+(\Omega t)^2 \right)^{\frac{s}{2}}} \sin \left( s \arctan\left(\Omega t\right) \right),
\end{equation}
with $\Gamma(s)$ the Euler gamma function,
which can be expressed in the equivalent, but more compact form, see Appendix \ref{app:a},
\begin{equation}\label{dephasingOK}
\mathcal{D}_s(t) = \lambda\Omega\Gamma(s)\frac{Im\left[(1+i\Omega t)^s\right]}{\left( 1+(\Omega t)^2 \right)^s}. 
\end{equation}
Correspondingly, the decoherence function can be written as
\begin{equation}\label{gammaOK}
\gamma_s(t) = \exp \left[ -\lambda\Gamma(s-1)\left(1-\frac{Re[(1+i\Omega t)^{s-1}]}{(1+(\Omega t)^2)^{s-1}}\right)\right].
\end{equation}

\begin{figure}[!ht]
{\bf (a)}
\\
\includegraphics[scale=0.55]{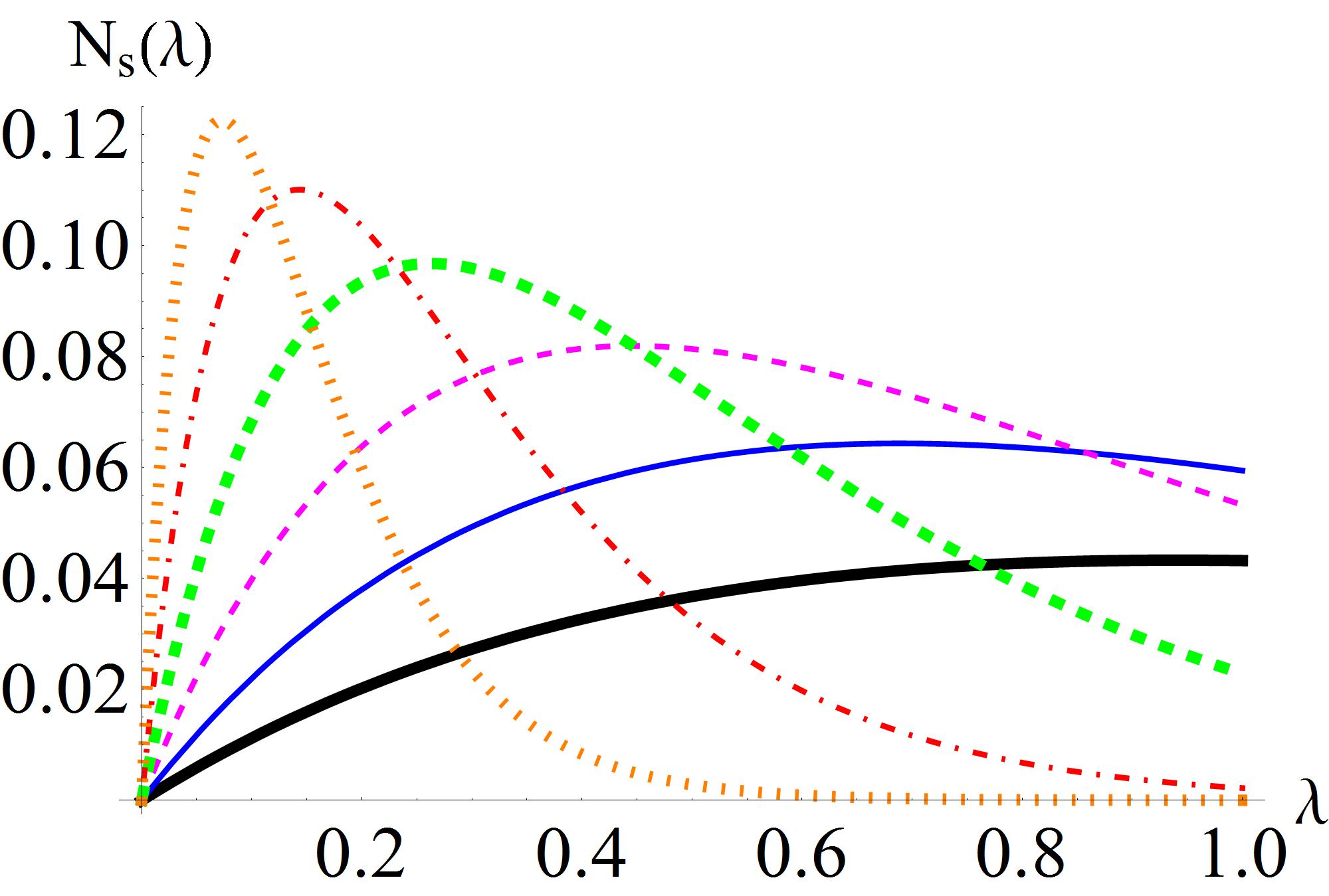}
\vspace{.5truecm}
\\
{\bf (b)}
\\
\includegraphics[scale=0.5]{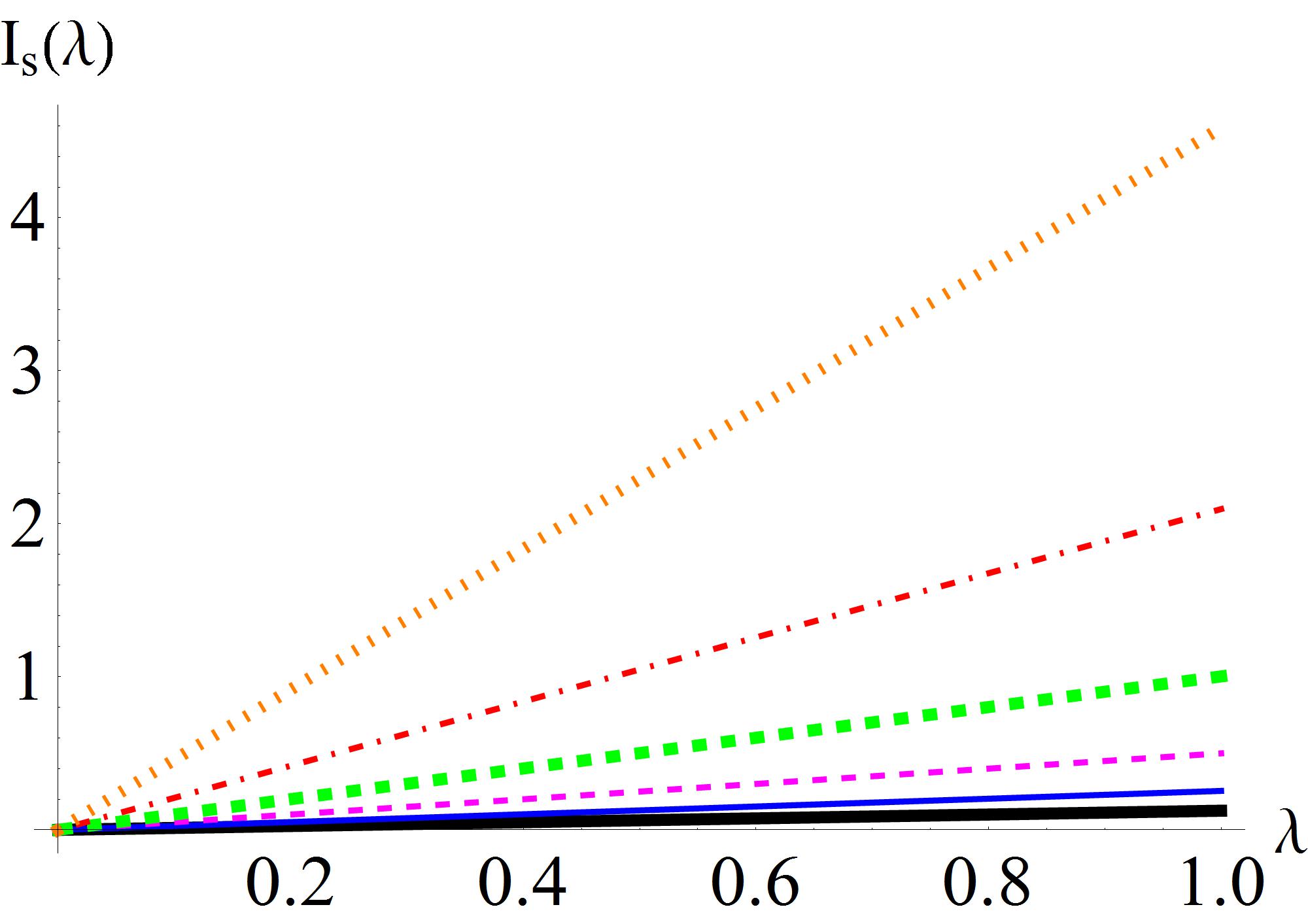}
\caption{(Color online) {\bf (a)}  BLP measure of non-Markovianity $\mathscr{N}_s(\lambda)$, see Eq.(\ref{eq:nmblp}), and {\bf (b)}  RHP measure of non-Markovianity $\mathscr{I}_s(\lambda)$, see Eq.(\ref{eq:nmrhp}), as a function of the coupling strength $ \lambda $ for increasing values of the parameter $s$. In both panels the curves are evaluated for $ s=3 $ (black thick solid line), $ s=3.5 $ (blue solid line), $ s=4 $ (magenta dashed line), $ s=4.5 $ (green dashed thick line), $ s=5 $ (red dot-dashed line) and $ s=5.5 $ (orange dotted line).}
\label{fig:nm}
\end{figure}

As before, let $\Omega_{+}$ be the union of the time intervals for which $\mathcal{D}(t) < 0$, i.e., equivalently, $|\gamma(t)|$ increases.
The number of solutions of the equation $ \mathcal{D}(t)=0 $ grows with the parameter $ s $: 
for $ s=1,2 $ the dephasing function is always strictly positive,
while for $ s=3 $ and  $ s=4 $ there is one zero at
$ t^*_3 = \frac{\sqrt{3}}{\Omega} $ and $ t^*_4 =
\frac{1}{\Omega} $  respectively.
Indeed, if the number of zeros is odd, $\mathcal{D}(t)$ is negative from its last zero to infinity,
while if the number of zeros is even, it approaches zero asymptotically from above.
As a consequence, the two measures of non-Markovianity are equal to zero for $s=1,2$ and, to give an example,
one has for $s=3$
\begin{eqnarray}
\mathscr{N}_3(\lambda)&=& \lim_{t\rightarrow \infty} |\gamma(t)| - |\gamma(t^*_3)| = e^{-\lambda }-e^{-\frac{9}{8} \lambda} \nonumber\\
\mathscr{I}_3(\lambda)&=&  \lim_{t\rightarrow \infty} \ln |\gamma(t)| - \ln|\gamma(t^*_3)|= \frac{\lambda}{8},
\end{eqnarray}
and, analogously, for $ s=4 $
\begin{eqnarray}
\mathscr{N}_4(\lambda)=e^{-2 \lambda }-e^{-\frac{5}{2} \lambda}, \quad \qquad
\mathscr{I}_4(\lambda)= \frac{\lambda}{2}.
\end{eqnarray}
In Fig.~\ref{fig:nm} {\bf (a)} and {\bf (b)}, we report, respectively, the BLP and the RHP measures
of non-Markovianity as a function of $\lambda$, for different values of $s$.

\begin{figure}[!ht]
{\bf (a)}
\\
\includegraphics[scale=0.52]{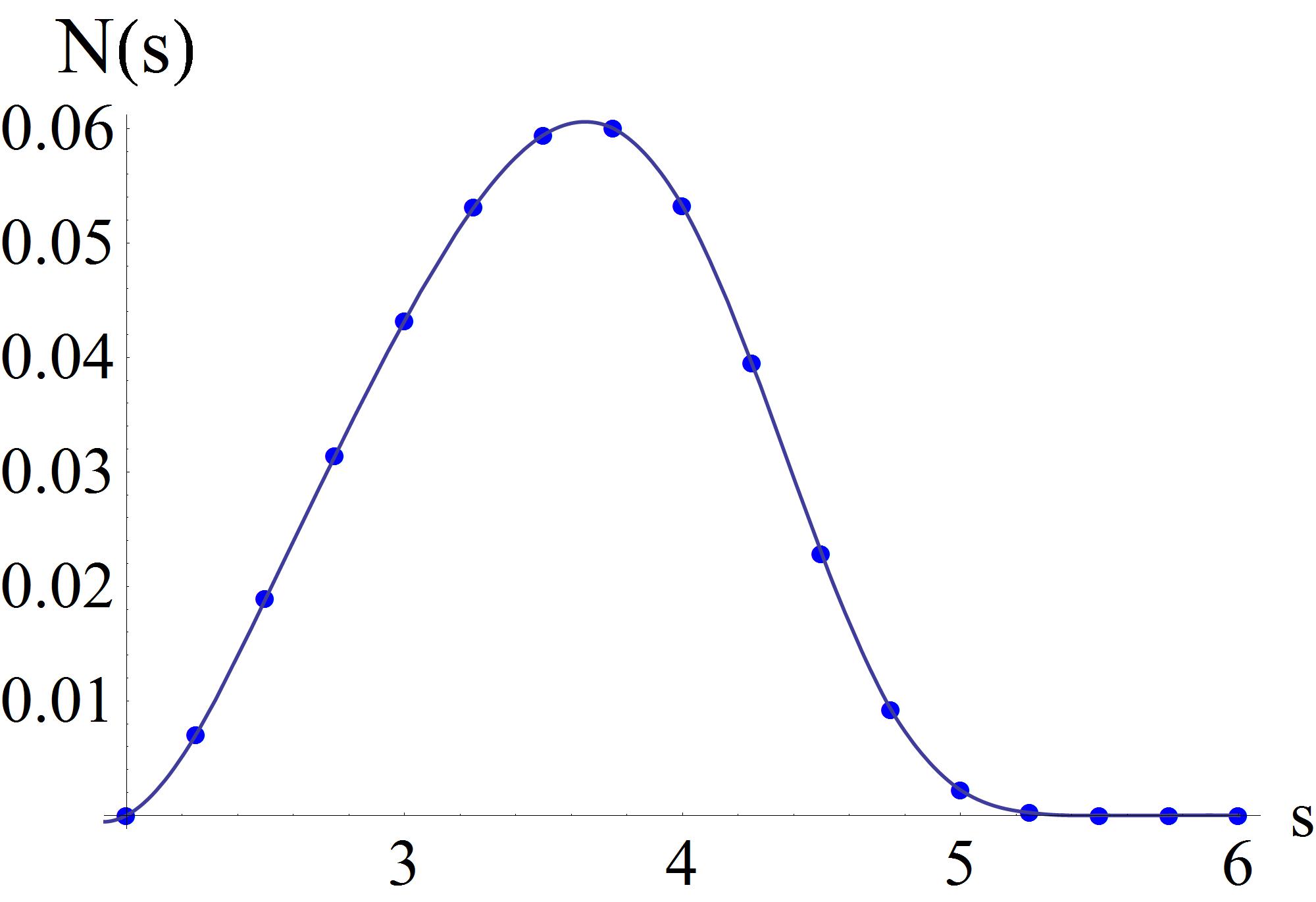}
\vspace{.5truecm}
\\
{\bf (b)}
\\
\includegraphics[scale=0.5]{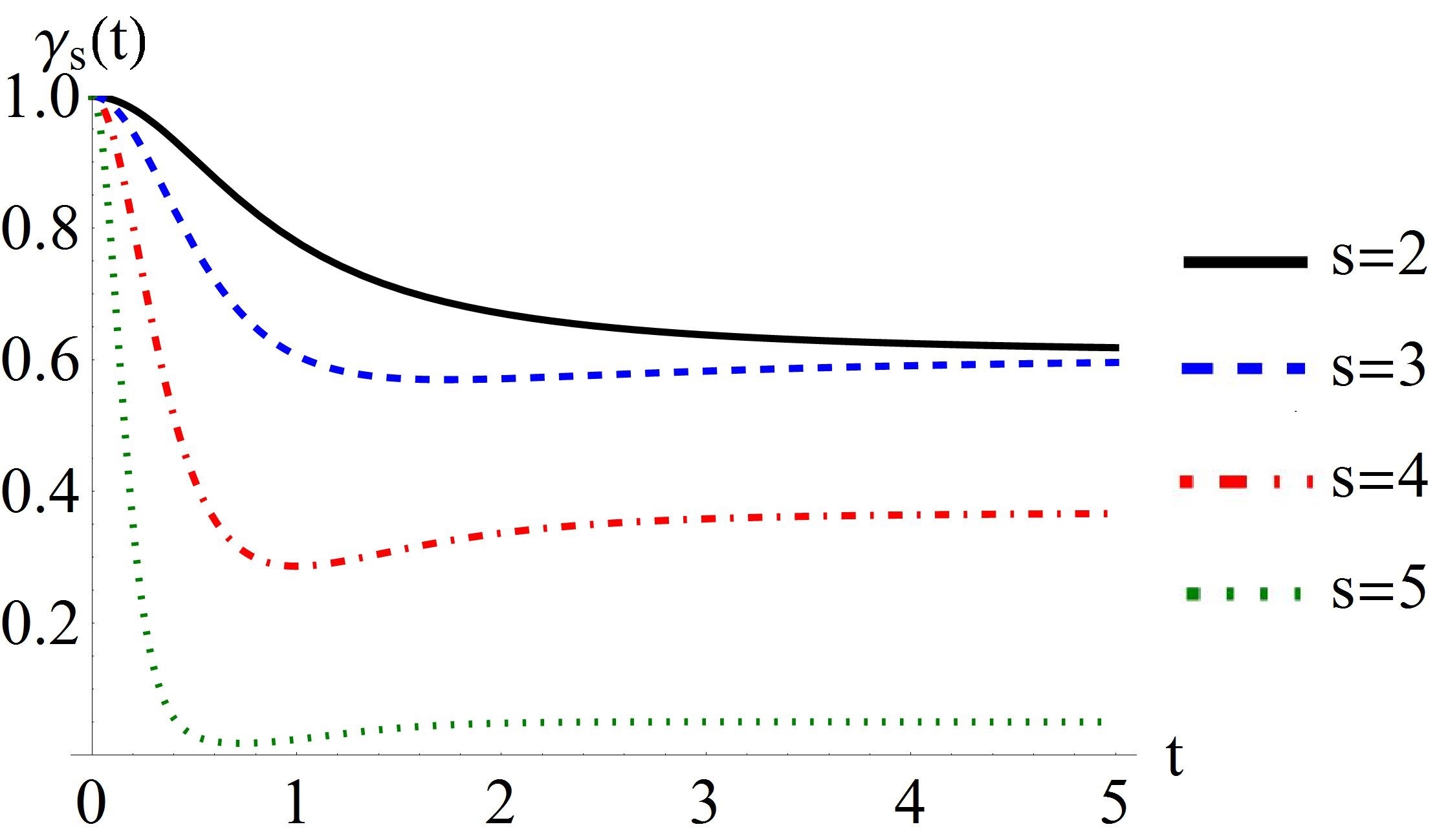}
\vspace{.5truecm}
\\
{\bf (c)}
\\
\includegraphics[scale=0.48]{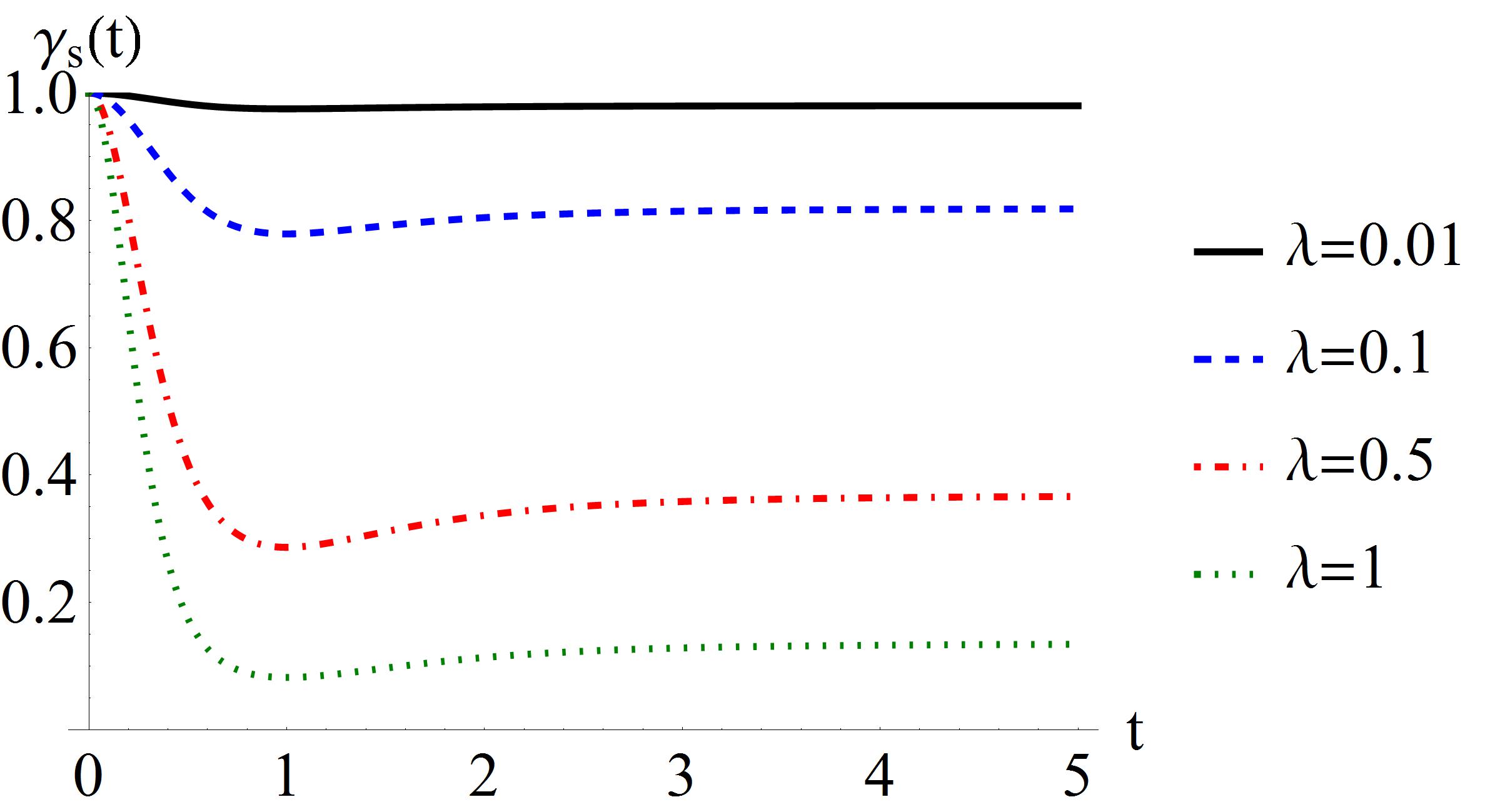}
\caption{{\bf (a)}  BLP measure of non-Markovianity
  $\mathscr{N}_s(\lambda)$, see Eq.(\ref{eq:nmblp}), as a function of
  the parameter $s$, for $\lambda = 1$. {\bf (b)} and {\bf (c)} Decoherence function $\gamma_s(t)$ as a function of time for $ \lambda = 0.5$  and different values of $s$ {\bf (b)}, and for $ s=4 $ and different values of $ \lambda $ {\bf (c)}. }
\label{fig:2}
\end{figure}

The behaviour of the two measures is clearly different. The RHP measure
is a monotonically increasing function of both $\lambda$ and $s$: the increase is linear with respect
to the former parameter and exponential with respect to the latter.
On the other hand,  for every fixed $s$, there is a critical value of the coupling strength $ \lambda^*(s) $, which
is smaller for increasing $s$, that separates two different regimes of the BLP measure:  for $ \lambda < \lambda^*(s) $, 
the non-Markovianity measure increases with the increase of the system-environment coupling,
while for $ \lambda > \lambda^*(s) $ it decreases with the increase of the coupling.
Analogously, there is a threshold value $s^*(\lambda)$ of the parameter $s$, which is higher for smaller values of $\lambda$,
such that the BLP measure increases for $s < s^{*}(\lambda)$ and decreases for $s>s^*(\lambda)$, see also Fig.~\ref{fig:2} {\bf (a)}.
Incidentally, the maximum value as a function of $\lambda$, 
$ \max_{\lambda} \mathscr{N}_s(\lambda) $, is a monotonically increasing function of the parameter $s$.
Indeed, the different behaviour of the non-Markovianity measures
traces back to their different functional dependence
of the decoherence function $\gamma_s(t)$, which is plotted in Fig.~\ref{fig:2} {\bf (b)} and {\bf (c)}
for different values of $s$ and $\lambda$.
One can see how $ \gamma_s(t) $ takes on smaller values within $[0,1]$ for growing values of $\lambda$, while its global minimum decreases with increasing $s$.
Now, while the BLP measure is fixed by the difference between the values of $\gamma_s(t)$ at the edges of the time intervals $[a_m,b_m]$ 
in which $\gamma_s(t)$ increases, see Eq. (\ref{eq:nmblp}), the RHP measure is fixed  by the ratio between the same values,  see Eq.(\ref{eq:nmrhp}).
Hence, as the coupling strength grows over the threshold $\lambda^*(s)$ or the parameter $s$ overcomes the threshold $s^*(\lambda)$,
the difference between $b_m$ and $a_m$ is increasingly smaller, 
and therefore $\mathscr{N}_s(\lambda)$ is so.
However, the ratio between $b_m$ and $a_m$ always increases with $\lambda$ and $s$, as witnessed by the
corresponding monotonic increase of $\mathscr{I}_s(\lambda)$.

\subsection{Validity of regression hypothesis}

\subsubsection{Exact expression versus quantum regression theorem}

The exact unitary evolution, Eq.(\ref{eq:EvolutionOperator}), directly provides us with
the average values, as well as the two-time correlation functions of the observables
of the system. In view of the comparison with the description given by the quantum regression theorem, see Sec.~\ref{sec:qrt},
let us focus on the basis of linear operators on $\mathbb{C}^2$, orthonormal with respect to the Hilbert-Schmidt
scalar product, given by $\left\{\Id/\sqrt{2}, \sigma_-, \sigma_+ ,\sigma_z/\sqrt{2}\right\}$.
Indeed, the first and the last element of the basis are constant of motion, see Eq.(\ref{rho_S(t)matrix}),
while the mean values of $\sigma_-$ and $\sigma_+$ evolve according to, respectively,
\begin{equation}\label{eq:meann}
\mean{\sigma_-(t)} = \gamma(t)e^{-i\omega_s t}  \mean{\sigma_- (0)}
\end{equation}
and the complex conjugate relation.
In a similar way, all the two-time correlation functions involving $\Id/\sqrt{2}$ or $\sigma_z/\sqrt{2}$
satisfy the condition of the quantum regression theorem in a trivial way,
as at most one operator within the two-time correlation function actually evolves.
The only non-trivial expressions are thus the following:
\begin{align}\label{eq:exact}
\mean{\sigma_-(t_2) \sigma_+(t_1)}\!&=\!e^{\scriptscriptstyle{-i\omega_s (t_2-t_1)}} \gamma(t_2,t_1) e^{\scriptscriptstyle{i \phi(t_2,t_1)}} \mean{\left(\sigma_-\sigma_+\right)(t_1)} \notag\\
\mean{\sigma_+(t_2) \sigma_-(t_1)}\!&=\!e^{\scriptscriptstyle{i\omega_s (t_2-t_1)}} \gamma^*(t_2,t_1) e^{\scriptscriptstyle{i \phi(t_2,t_1)}} \mean{\left(\sigma_+\sigma_-\right) (t_1)}
\end{align}
where 
\begin{equation}\label{gammatwopoints}
\gamma(t_2,t_1) =  \tmop{Tr}_E{\rho_E \prod_k \Delta(\alpha_k(t_2)-\alpha_k(t_1))} 
\end{equation} 
and
\begin{equation} \label{phi}
\phi(t_2,t_1) =  \sum_k Im\left[\alpha_k^*(t_2) \alpha_k(t_1)\right] .
\end{equation}
Here, {to derive \eqref{eq:exact}} we used
the properties of the displacement operator \cite{Ferraro2005}
\[
\Delta(\alpha) \Delta(\beta) = \Delta(\alpha+\beta)e^{i Im(\alpha \beta^*)}, \quad \Delta^{\dag}(\alpha) = \Delta(-\alpha),
\]
and the equality $\mean{\left(\sigma_+ \sigma_-\right) (t)} =  \mean{\sigma_+ \sigma_- }$.

We can now obtain the corresponding two-time correlation functions
as predicted by the quantum regression theorem.
By Eq.(\ref{eq:meann}), one has 
\begin{equation}\label{eq:meannd}
\frac{\mathd}{\mathd t}\mean{\sigma_-(t)} =  \left(\frac{\mathd \gamma(t)/\mathd t}{\gamma(t)}- i \omega_s \right) \mean{\sigma_- (t)}
\end{equation}
and the complex conjugate relation for $\mean{\sigma_+(t)}$.
The specific choice of the operator basis has lead us to a diagonal matrix $G$
in Eq.(\ref{lindiffeq}). Hence, one has immediately
\begin{eqnarray}
\mean{\sigma_-(t_2) \sigma_+(t_1)}_{qrt} =e^{-i\omega_s (t_2-t_1)} \frac{\gamma(t_2)}{\gamma(t_1)} \mean{\sigma_- (t_1)\sigma_+ (t_1)} \nonumber\\
\mean{\sigma_+(t_2) \sigma_-(t_1)}_{qrt} = e^{i\omega_s (t_2-t_1)} \frac{\gamma^*(t_2)}{\gamma^*(t_1)} \mean{\sigma_+ (t_1)\sigma_- (t_1)}. \nonumber\\ \label{eq:qrt}
\end{eqnarray}
The quantum regression theorem will be generally violated within this model, compare Eq.(\ref{eq:exact}) and (\ref{eq:qrt}).
We quantify such a violation by means of the figure of merit introduced in Eq.(\ref{eq:figuremerit}),
which for the couple of operators $ \sigma_- $ and $ \sigma_+ $ reads
\begin{align}\label{Zmp}
Z &= \left| 1 - \frac{\mean{\sigma_-(t_2)\sigma_+(t_1)}_{qrt}}{\mean{\sigma_-(t_2)\sigma_+(t_1)}} \right| \notag\\
&= \left| 1 - \frac{\gamma(t_2)}{\gamma(t_1)\gamma(t_2,t_1)e^{i\phi(t_2,t_1)}} \right|. 
\end{align}

\subsubsection{Quantitative analysis of the violations of the quantum regression theorem}

The expressions of the previous paragraph hold for generic initial 
state of the bath and spectral density. Now, we come back
to the specific choice of an initial thermal bath. The results in Eq.(\ref{eq:qrt})
are in this case in agreement with those found in \cite{Tai2010PRA}, 
where the two-time correlation functions have been evaluated
focusing on a spectral density as in Eq.(\ref{spectralS}) with $s=1$,
while keeping a generic temperature of the bath. Instead, we will focus on the case $T=0$
and  maintain a generic value of $s$ in order to compare the behaviour of the two-time
correlation functions with the measures of non-Markovianity.

\begin{figure}[!ht]
{\bf (a)}
\\
\includegraphics[scale=0.35]{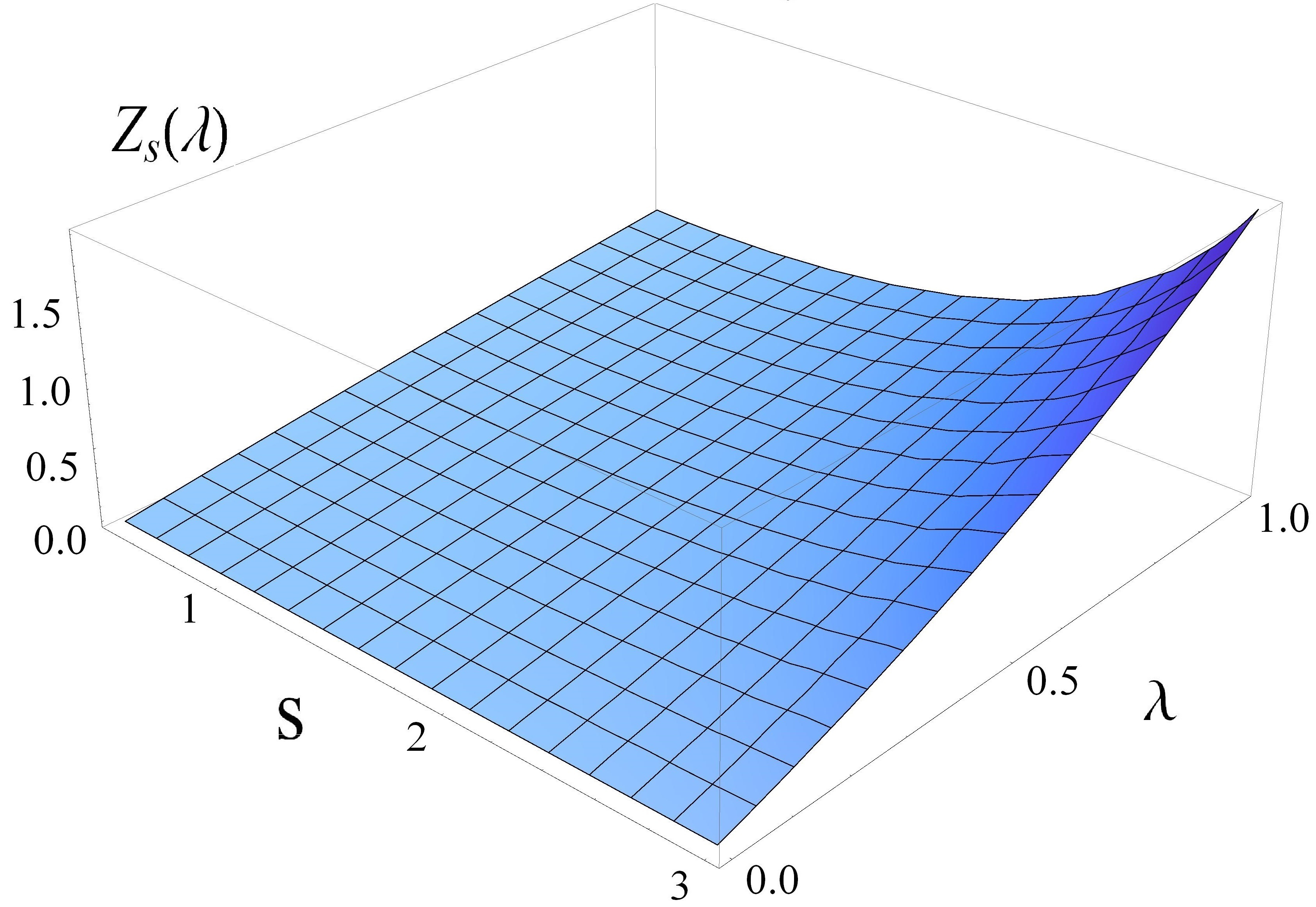}
\vspace{.5truecm}
\\
{\bf (b)}
\\
\includegraphics[scale=0.55]{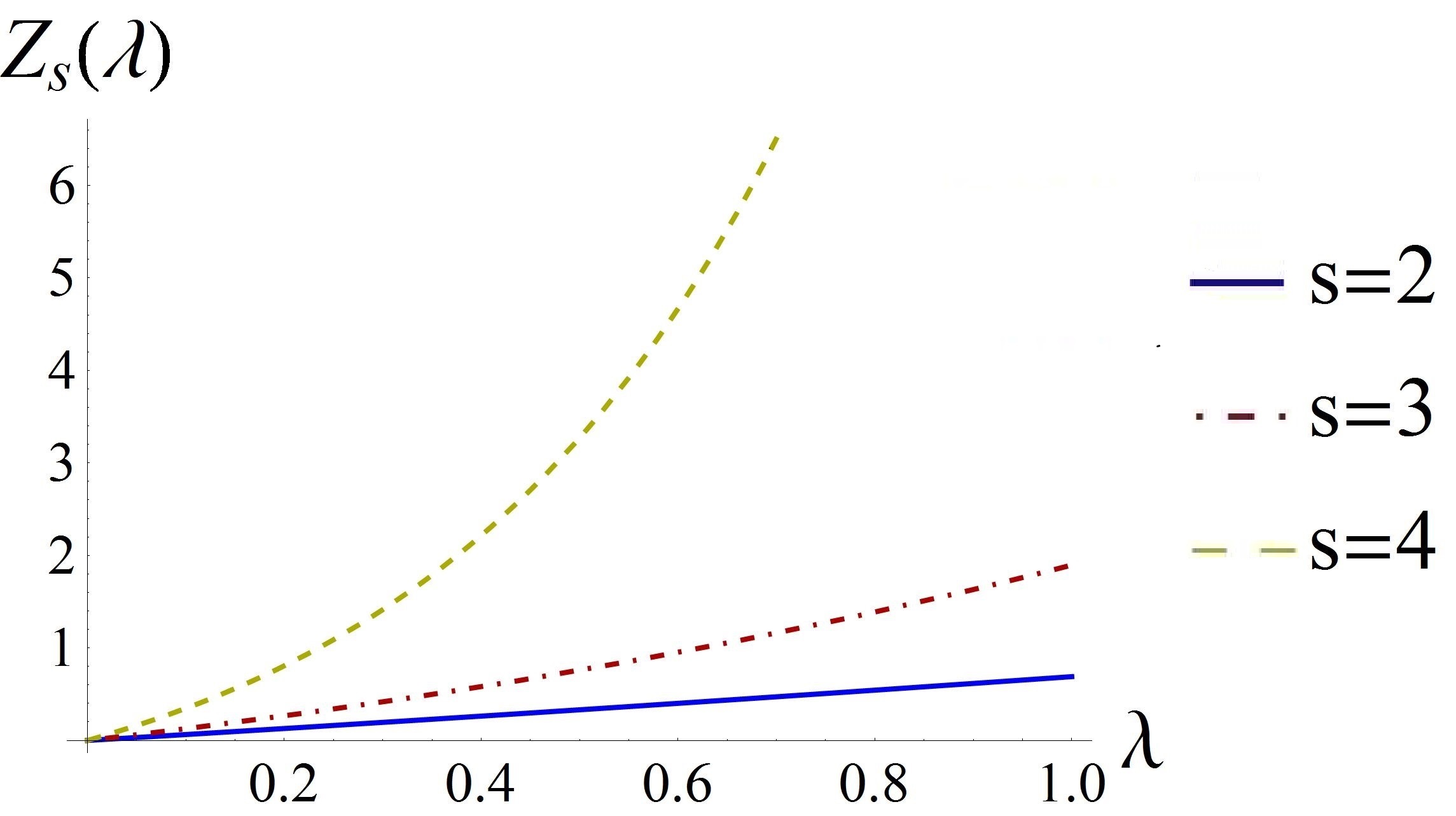}
\caption{{\bf (a)} $ Z_s(\lambda) $ as a function of the parameter $s$ and of the coupling strength $ \lambda $, see Eq.(\ref{eq:zslambda}), for $ \Omega t_1 = 1 $ and $ \Omega t_2 = 2 $.
{\bf (b)} Section of {\bf (a)} for $s=2,3,4$.}
\label{fig:Zs}
\end{figure}

First, note that by using the definition of the displacement operator as well as Eq.(\ref{eq:alpha}),
one can show the general identity
\begin{equation}\label{eq:rel}
\Delta(\alpha_k(t_2)-\alpha_k(t_1)) =  \Delta\left(\alpha_k(t_2-t_1) e^{i \omega_k t_1 }\right).
\end{equation}
But then, since for a thermal state
$\tmop{Tr}_E{\Delta(\alpha)\rho_E}$ is a function of $|\alpha|$ only \cite{Breuer2002}, Eq.(\ref{eq:rel})
implies 
\begin{equation}\label{eq:tt}
\gamma(t_2,t_1) = \gamma(t_2-t_1),
\end{equation}
see Eqs.(\ref{gammatwopoints}) and (\ref{decoherentfunction}).
In addition we have in the continuum limit, see Eq.(\ref{phi}),
\[
\phi(t_2,t_1)\!=\!\int\! \mathd \omega\!\frac{J(\omega)}{\omega^2}\left[\sin( \omega t_2)\!-\!\sin(\omega t_1)\!-\!\sin(\omega (t_2-t_1)) \right]
\]
so that, for $J(\omega)$ as in Eq.(\ref{spectralS}) and using
Eq.(\ref{eq:Dephasing}) in the zero temperature limit, we get
\begin{equation}\label{eq:phiz}
\phi_s(t_2,t_1) = ( \mathcal{D}_{s-1}(t_2) - \mathcal{D}_{s-1}(t_1)-\mathcal{D}_{s-1}(t_2-t_1))/\Omega.
\end{equation}
The identities in Eqs.(\ref{dephasingOK}) and (\ref{gammaOK}),
along with Eqs. ({\ref{eq:tt}}) and (\ref{eq:phiz}), finally 
provide us with the explicit expression of the estimator for
the violations of the quantum regression theorem, see Eq.(\ref{Zmp}),
\begin{multline}\label{eq:zslambda}
Z_s(\lambda) = \left| 1- \exp\left[ \lambda\Gamma(s-1) \left[1 - (1+i \Omega(t_2-t_1))^{1-s}\right.\right.\right.\\
\left.\left.\left.-(1+i \Omega t_1)^{1-s}+(1+i \Omega t_2)^{1-s} \right) \right]\right|,
\end{multline}
whose behaviour as a function of $\lambda$ and $s$ is shown in Fig.~\ref{fig:Zs} {\bf (a)} and {\bf (b)}.
The violation of the quantum regression theorem monotonically increases with increasing values of both the
coupling strength $\lambda$ and the  parameter $s$. This behaviour 
is clearly in agreement with that of the RHP measure of non-Markovianity,
see Sec. \ref{sec:zte} and in particular Fig.~\ref{fig:nm}. 
From a quantitative point of view there is, however, some difference as
the estimator
$Z_s(\lambda)$, {at variance with the RHP measure}, grows linearly with $\lambda$ only for small values of $s$, while it
growths faster for $s > 3$; compare with Fig.~\ref{fig:nm} {\bf(b)}. 
In any case, the RHP measure appears to be more directly related with the strength of
the violation to the quantum regression theorem, as compared with the BLP measure.
This can be traced back to the different influence of the system-environment correlations on the two measures.
As we recalled in Sec. \ref{sec:qrt}, the hypothesis that the
state of the total system at any time $t$ is well approximated
by the product state between the state of the open
system and the initial state of the environment, see Eq.(\ref{eq:prodt}),
lies at the basis of the quantum regression theorem. This
hypothesis is expected to hold in the weak coupling regime,
while for an increasing value of $\lambda$, the interaction will build stronger system-environment
correlations, leading to a strong violation of the quantum regression theorem. 
The establishment of correlations between the system and the environment
due to the interaction plays a significant role also in the subsequent
presence of memory effects in the dynamics of the open system \cite{Laine2010bPRA,Mazzola2012PRA,Smirne2013PRA}.
Indeed, different signatures of the memory effects can be affected by
system-environment correlations in different ways. 
In particular, the CP-divisibility of the dynamical maps appears to be a more fragile property than
the contractivity of the trace distance and therefore it is more
sensitive to the violations of the quantum regression theorem.
Furthermore, it is worth noting that the estimator
$Z_s(\lambda)$ steadily increases with the coupling strength $\lambda$ even for values
of $s$ such that the corresponding reduced dynamics is Markovian according to either definitions. The validity of the quantum regression theorem
calls {therefore} for stricter conditions than the Markovianity of quantum dynamics.

\section{Photonic realization of dephasing interaction}
\label{sec:modellofotoni}

In the pure dephasing spin-boson model, there is no regime in which the quantum regression theorem
is strictly satisfied, apart from the trivial case $\lambda = 0$.
In addition, we have shown that the strength of the violations of this theorem
has the same qualitative behaviour of the RHP non-Markovianity measure,
as they increase with both $\lambda$ and the parameter $s$.
In this section, we take into account a different pure dephasing model,
which allows us to deepen our analysis on the relationship between
the quantum regression theorem and the Markovianity of the reduced-system
dynamics. In particular, we show that in general these two notions should be considered
as different since the quantum regression theorem may be strongly violated,
even if the open system's dynamics is Markovian, irrespective of the exploited definition.

\subsection{The model}

Let us deal with the pure-dephasing interaction considered in Ref.~\cite{Liu2011NAT}.
The open system here is represented by the polarization degrees of freedom of a photon
generated by spontaneous parametric down conversion,
while the environment consists in the corresponding frequency degrees of freedom.
The overall unitary evolution, which is realized via a quartz
plate {that couples} the polarization and frequency degrees of freedom,
can be described as
\begin{equation}\label{eq:ut}
U(t) \ket{j, \omega} = e^{i n_j \omega t} \ket{j, \omega} \qquad j=0,1,
\end{equation}
where
$ \ket{0} \equiv \ket{H}$ and $\ket{1} \equiv \ket{V}$ are the two polarization states  (horizontal and vertical), with refractive indexes, respectively, 
$ n_0 \equiv n_H$ and $ n_1 \equiv n_V $, while $\ket{\omega}$ is the environmental state with frequency $\omega$.
If we consider an initial product state, see Eq.(\ref{eq:prod}),
with a pure environmental state $ \rho_E= \ket{\Psi_E} \bra{\Psi_E} $, where
\begin{equation}\label{eq:psie}
\ket{\Psi_E} = \int \, d\omega\, f(\omega) \ket{\omega},
\end{equation} 
we readily obtain that the reduced dynamics is given by 
Eq.(\ref{rho_S(t)matrix}).
Again, we are in the presence of a pure dephasing dynamics,
the only difference being the decoherence function, which now reads
\begin{equation}\label{decoherencenature}
\gamma(t) = \int \, d\omega\, \left| f(\omega) \right|^2 e^{i \Delta n \omega t},
\end{equation}
with $ \Delta n \equiv  n_1-n_0 $.
For the rest, the results of Secs. \ref{sec:tm} and \ref{sec:nmm}
directly apply also to this model: the master equation is given by Eq.(\ref{eq:me}),
with $\epsilon(t)$ and $\mathcal{D}(t)$ as in, respectively, Eq.(\ref{eq:eps}) (for $\omega_s=0$) and Eq.(\ref{eq:dt}),
while the non-Markovianity measures are as in Eq.(\ref{eq:nmblp}) and Eq.(\ref{eq:nmrhp}).
Analogously, the two-time correlation functions are given by Eq. (\ref{eq:exact})
with 
\begin{equation}
\gamma(t_2,t_1) = \gamma(t_2-t_1) \qquad
\phi(t_2,t_1)= 0,
\end{equation}
while the application of the quantum regression theorem leads to the expressions in Eq.(\ref{eq:qrt}) (with $\omega_s=0$).
Hence, the violations of the quantum regression theorem can
be quantified by
\begin{equation}\label{Znature}
\!\!\!
Z = \left| 1 - \frac{\mean{\sigma_-(t_2)\sigma_+(t_1)}_{qrt}}{\mean{\sigma_-(t_2)\sigma_+(t_1)}}\right| = \left| 1 - \frac{\gamma(t_2)}{\gamma(t_1)\gamma(t_2-t_1)} \right|. 
\end{equation}

\subsection{Lorentzian frequency distributions}

\subsubsection{Semigroup dynamics}

Despite its great simplicity, this model
allows to describe the transition between Markovian and non-Markovian dynamics
in concrete experimental settings  \cite{Liu2011NAT,Tang2012EPL}.
Different dynamics are
obtained for different choices of the initial environmental state, see Eq.(\ref{eq:prod})
and the related discussion, i.e., for different initial frequency distributions, see Eq.(\ref{eq:psie}).
The latter can be experimentally set, e.g., by properly rotating a Fabry-P{\'e}rot cavity, through 
which a beam of photons generated by spontaneous
parametric down conversion passes \cite{Liu2011NAT}.
A natural benchmark is represented by the Lorentzian distribution
\begin{equation}
\left|f(\omega)\right|^2  = \frac{\delta \omega}{\pi\left[(\omega-\omega_0)^2+(\delta\omega)^2\right]},
\end{equation}
where $ \delta\omega $ is the width of the distribution and $ \omega_0 $ its central frequency,
as this provides a reduced semigroup dynamics \cite{Smirne2013PRA}.
The decoherence function, which is given by the Fourier transform of the frequency distribution, see  Eq.(\ref{decoherencenature}), is in fact
\begin{equation}
\gamma(t) = e^{-\Delta n (\delta\omega - i\omega_0) t}.
\end{equation}
Thus, replacing this expression in Eqs. (\ref{eq:eps}) and (\ref{eq:dt}),
one obtains a Lindblad equation, given by Eq. (\ref{eq:me}) with $\epsilon(t)=- \Delta n \, \omega_0$ and $\mathcal{D}(t) =\Delta n\, \delta \omega$.
In addition,
 $\gamma(t_2-t_1) = \gamma(t_2)/\gamma(t_1)$ and hence, 
 as one can immediately see by Eq.(\ref{Znature}),
$Z=0$. For this model, 
as long as the reduced dynamics is determined by a completely positive semigroup,
the quantum regression theorem is strictly valid.
Let us emphasize, that this is the case even if the total state is not a product state at any time $t$.
For example
if the initial state of the open system is the pure state $\ket{\psi_S}= \alpha \ket{H} + \beta \ket{V}$, with $|\alpha|^2+|\beta|^2=1$,
the total state at time $t$ is
\begin{equation}
\ket{\psi_{SE}(t)} = \int \mathd \omega f(\omega) (\alpha e^{i n_{H} \omega t}\ket{H, \omega} +\beta e^{i n_{V} \omega t}\ket{V, \omega}).
\end{equation}
This is an entangled state, of course unless $\alpha=0$ or $\beta=0$;
nevertheless, the quantum regression theorem does hold. 
This clearly shows that for the quantum regression theorem, 
as for the semigroup description of the dynamics \cite{Rivas2010NJP,Mazzola2012PRA,Smirne2013PRA},
the approximation encoded in Eq.(\ref{eq:prodt})
should be considered as an effective description of the total state, which can be
very different from its actual form, even when the theorem is valid.

\subsubsection{Time-inhomogeneous Markovian and non-Markovian dynamics}

\begin{figure}[!ht]
{\bf (a)}
\\
\includegraphics[scale=0.5]{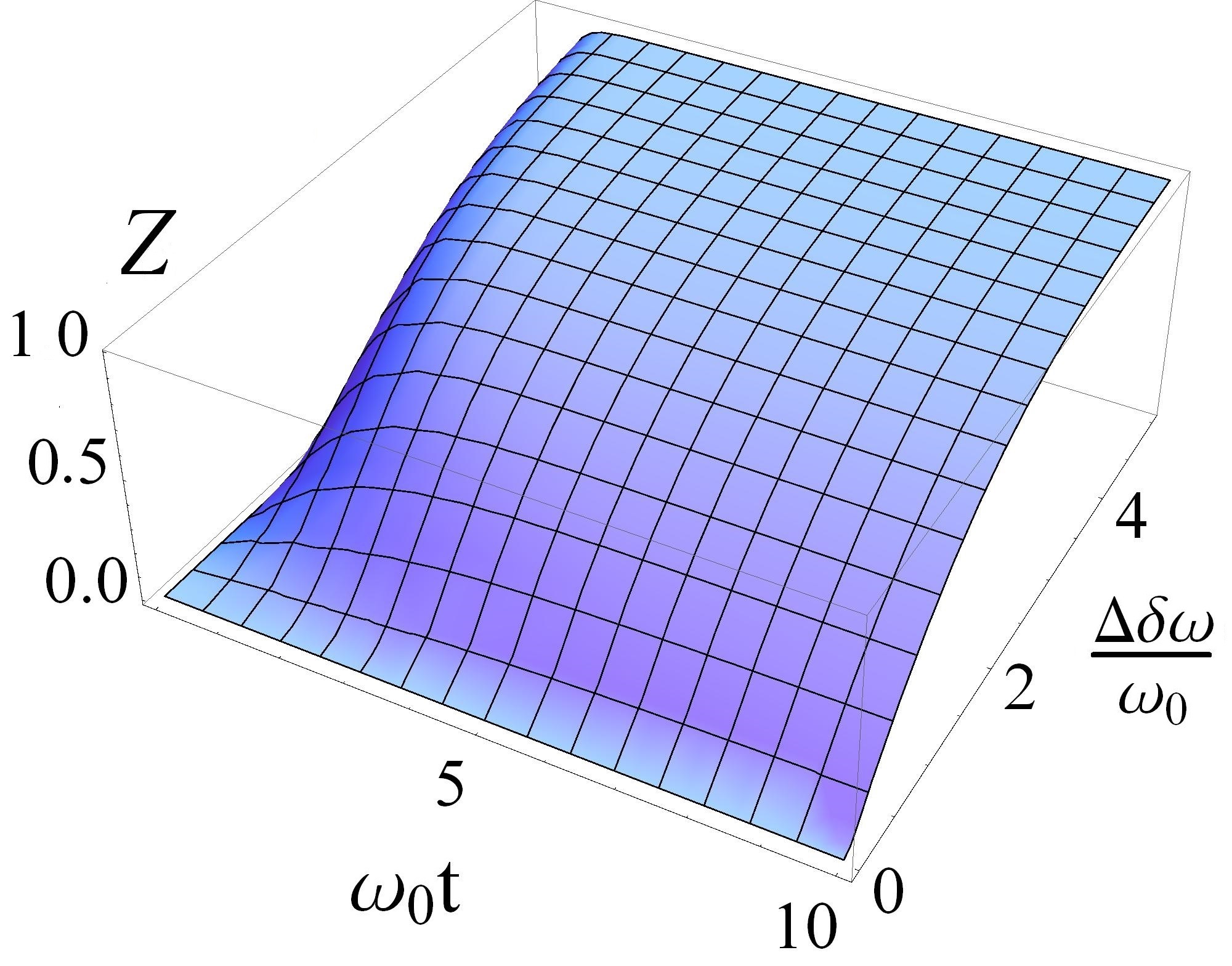}
\vspace{.5truecm}
\\
{\bf (b)}
\\
\includegraphics[scale=0.5]{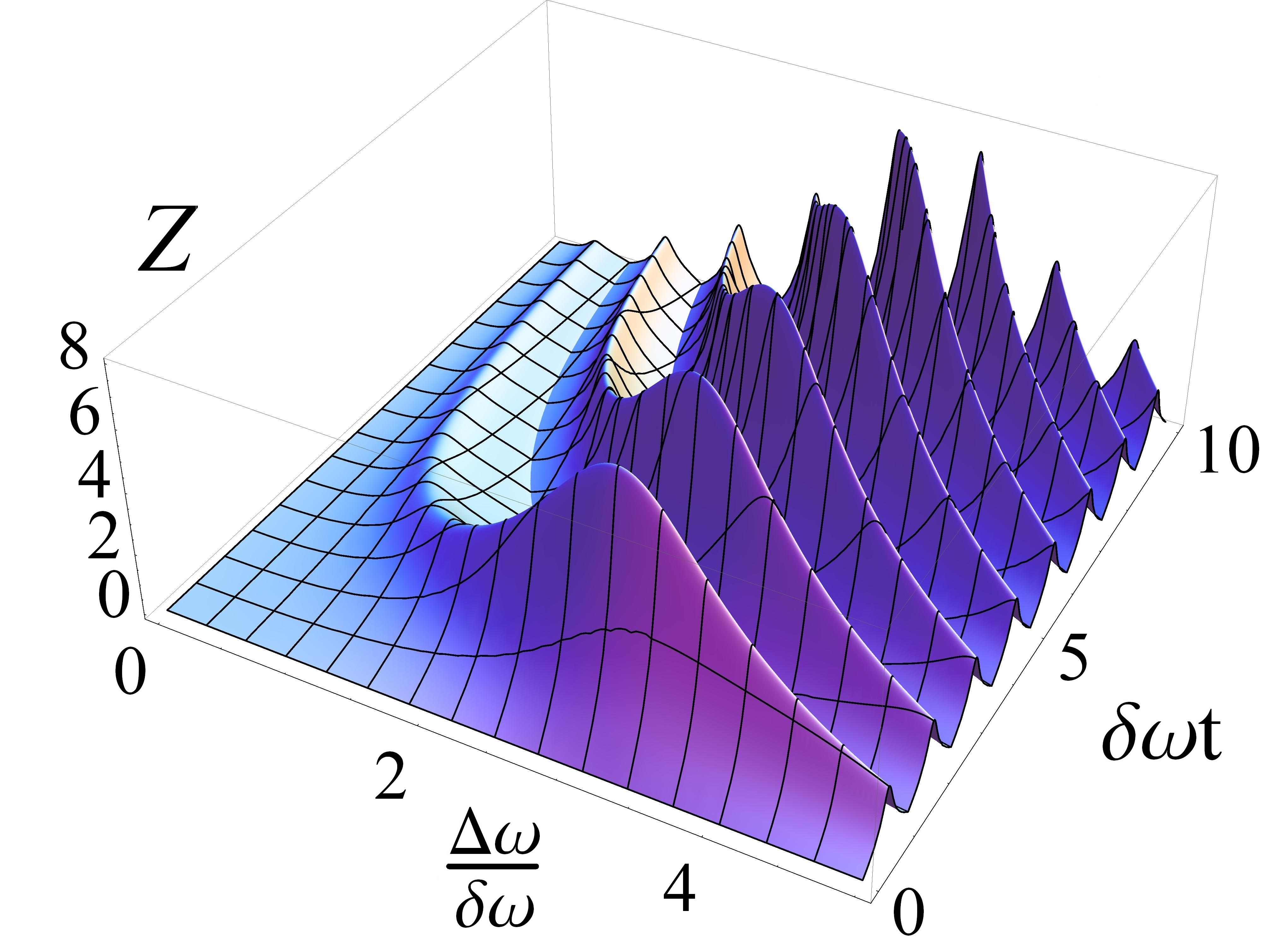}
\caption{Violation of the quantum regression theorem, as quantified by the estimator $Z$ in Eq.(\ref{Znature}) {\bf (a)} in the time-inhomogeneous
Markovian case, $\omega_{0,1} = \omega_{0,2} = \omega_0$, as a function of $\Delta\delta\omega = \delta \omega_1 - \delta \omega_2$ and $\omega_0\tau =\omega_0(t_2-t_1)$, for $\omega_{0} t_1 = 1$ and $ r=1 $; {\bf (b)} in the non-Markovian case, $\delta \omega_1 = \delta \omega_2 = \delta \omega$, as a function of $\Delta\omega_{0} = \omega_{0,1} - \omega_{0,2}$ and $\delta\omega\,\tau$, for $\delta\omega\, t_1 = 1$ and $ r=2 $; in all the panels $ \Delta n = 1 $.}
\label{fig:4}
\end{figure}

Now, we consider a more general class of frequency distributions; namely,
the linear combination of two Lorentzian distributions,
\begin{equation}\label{eq:lld}
\left|f(\omega)\right|^2  = \sum_{j=1,2} \frac{A_j\delta\omega_j}{\pi\left[(\omega-\omega_{0,j})^2+(\delta\omega_j)^2\right]},
\end{equation}
with $A_1+A_2=1$.
The decoherence function \eqref{decoherencenature} is in this case
\begin{equation}\label{eq:gga}
\gamma(t) = \frac{ e^{-\Delta n (\delta\omega_1 - i\omega_{0,1}) t} + r  e^{-\Delta n (\delta\omega_2 - i\omega_{0,2}) t}}{1+r},
\end{equation}
with $ r \equiv \frac{A_2}{A_1} $, while the estimator of the violations of the quantum regression theorem, see Eq.(\ref{Znature}),
can be written as a function of the difference between the central
frequencies, $\Delta\omega  = \omega_{0,1}-\omega_{0,2}$, as
well as of the difference between the corresponding widths, $\Delta\delta\omega= \delta \omega_{1} - \delta \omega_{2}$.
If we assume that the two central frequencies are equal, $ \omega_{0,1} = \omega_{0,2}= \omega_0 $,
the evolution of the two-level statistical operator is fixed by a time-local master equation as in Eq.(\ref{eq:me}),
with $\epsilon(t) = - \Delta n \, \omega_0$ and
\begin{equation}
\mathcal{D}(t) = \Delta n \frac{\delta \omega_1  e^{-\Delta n \delta\omega_1 t} + r\,\delta \omega_2  e^{-\Delta n \delta\omega_2 t} }{ e^{-\Delta n \delta\omega_1 t} + r\,e^{-\Delta n \delta\omega_2 t}}.
\end{equation}
The latter is a positive function of time:
the reduced dynamics is CP-divisible, see Sec.~\ref{sec:nmrhp},
and hence it is Markovian with respect to both the BLP and RHP definitions. Indeed,
now we are in the presence of a time-inhomogeneous Markovian dynamics.
Nevertheless, as $\gamma(t_2-t_1) \neq \gamma(t_2)/\gamma(t_1)$ the quantum regression theorem is violated, see Eq.(\ref{Znature}).
This is explicitly
shown in 
Fig.~\ref{fig:4} {\bf (a)}, where $Z$ is plotted as a function of $\Delta\delta\omega = \delta \omega_1 - \delta \omega_2$ and $\Delta n \tau$, with $\tau = t_2-t_1$.
With growing difference between the two widths, as well as the length of the time interval,
the deviations from
the quantum regression theorem are increasingly strong, up to a saturation value of the estimator $Z$.
Contrary to the semigroup case, here, even if the dynamics is Markovian according to both definitions,
the actual behaviour of the two-time correlation functions cannot be reconstructed by the evolution of the mean values.

Finally, let us consider a frequency distribution as in Eq. \eqref{eq:lld},
but now with $\delta\omega_1 = \delta\omega_2 = \delta\omega$ and $\omega_{0,1} \neq \omega_{0,2}$.
This frequency distribution has two peaks and the resulting reduced
dynamics is non-Markovian \cite{Liu2011NAT,Smirne2013PRA}.
{In this case the BLP non-Markovianity measure \eqref{eq:nblp}} increases with the increasing of the distance between the two peaks,
while the estimator $Z$ grows for small values of the distance and then it exhibits
an oscillating behaviour, see Fig.~\ref{fig:4} {\bf (b)}. Indeed, for $\Delta\omega=0$ one recovers the semigroup dynamics
previously described and, accordingly, $Z$ goes to zero.
Summarizing,
by varying the distance between the two peaks, one obtains a transition from a Markovian (semigroup)
dynamics to a non-Markovian one and, correspondingly, the quantum regression theorem 
ceases to be satisfied and is even strongly violated. Nevertheless,
the qualitative behaviour of, respectively, the non-Markovianity of the reduced dynamics and 
the violation of the quantum regression theorem appear to be different.

\section{Conclusions}
\label{sec:ceo}
We have explored the relationship between two criteria for
Markovianity of a quantum dynamics, namely the CP-divisibility of the
quantum dynamical map and the behaviour in time of the trace distance
between two distinct initial states, and the validity of the quantum
regression theorem, which is a statement relating the behaviour in time
of the mean values and of the two-time correlation functions of system operators.
{The first open system considered
is a two-level system} affected by a bosonic environment
through a dephasing interaction. For a class of spectral densities
with exponential cut-off and power law behaviour at low frequencies we
have studied the onset of non-Markovianity as a function of the
coupling strength and of the power determining the low frequency
behaviour, further giving an exact expression for the corresponding
non-Markovianity measures. The deviation from the quantum regression
theorem has been estimated evaluating the relative error made in
replacing the exact {two-time} correlation function for the system
operators with the expression reconstructed
by the evolution of the corresponding mean values. 
It appears that the validity of the quantum regression theorem
{represents} a stronger requirement than Markovianity, according to
either criteria, which in this case coincide but quantify non-Markovianity
in a different way and exhibit distinct performances in their dependence
{on strength of the coupling and low frequency behaviour. We have further
considered an all-optical realization of a dephasing interaction, as recently
exploited for the experimental investigation of non-Markovianity, obtaining also in
this case, for different choices of the frequency distribution, significant violations to
the quantum regression theorem even in the presence of a Markovian
dynamics.}

These results suggest that indeed the recently introduced new
approaches to quantum non-Markovianity provide a weaker requirement
with respect to the classical notion of Markovian classical
process. Further and more stringent notion of Markovian quantum
dynamics can therefore be introduced, e.g. relying on validity of
the quantum regression theorem \cite{Logullo2014arxiv}. However, the usefulness of such criteria
will heavily depend on the possibility to verify their satisfaction
directly by means of experiments, as it is the case e.g. for the notion
of Markovianity based on trace distance, without asking for an
explicit exact knowledge of the dynamical equations.

\acknowledgments
The authors gratefully
acknowledge financial support by the EU projects COST
Action MP 1006 and NANOQUESTFIT.

\appendix
\section{Alternative expression of the dephasing function}\label{app:a}

Starting from Eq.~\eqref{dephasingT0}, namely
\begin{equation}
\mathcal{D}_s(t) = \frac{\lambda\Omega\Gamma(s)}{\left( 1+(\Omega t)^2 \right)^{\frac{s}{2}}} \sin \left( s \arctan\left(\Omega t\right) \right),
\end{equation}
and exploiting the identities
\begin{equation}
\sin\left( \arctan (x)\right) =\!\frac{x}{\sqrt{1+x^2}}\!\quad ,\cos\left( \arctan (x)\right) =\! \frac{1}{\sqrt{1+x^2}}
\end{equation}
together with
\begin{equation}
\sin\left( s x \right) = \sum_{k=0} \begin{pmatrix}s\\k\end{pmatrix} \left(\cos(x)\right)^k \left(\sin(x)\right)^{s-k}\sin\left(\frac{\pi}{2}(s-k)\right),
\end{equation}
we can come to the compact expression \eqref{dephasingOK}
\begin{align}
\mathcal{D}_s(t)
\!&=\! \frac{\lambda\Omega\Gamma(s)}{2i \left( 1+(\Omega t)^2 \right)^{{s}}}\!\left[\sum_{k=0} \begin{pmatrix}s\\k\end{pmatrix}\! \left(\Omega t\right)^{\scriptscriptstyle{s-k}}\! \left(i^{\scriptscriptstyle{s-k}}\! -\! (-i)^{\scriptscriptstyle{s-k}}\! \right)\right]\notag\\
&= \frac{\lambda\Omega\Gamma(s)}{2i\left( 1+(\Omega t)^2 \right)^s} \left[(1+i\Omega t)^s - (1-i\Omega t)^s\right] \notag\\
&=  \lambda\Omega\Gamma(s)\frac{Im\left[(1+i\Omega t)^s\right]}{\left( 1+(\Omega t)^2 \right)^s}.
\end{align}

\end{document}